\documentstyle[twoside,epsf]{article}

\catcode`\@=11
\long\def\@makefntext#1{
\protect\noindent \hbox to 3.2pt {\hskip-.9pt
$^{{\eightrm\@thefnmark}}$\hfil}#1\hfill}		

\def\@makefnmark{\hbox to 0pt{$^{\@thefnmark}$\hss}}	

\def\ps@myheadings{\let\@mkboth\@gobbletwo
\def\@oddhead{\hbox{}
\rightmark\hfil\eightrm\thepage}
\def\@oddfoot{}\def\@evenhead{\eightrm\thepage\hfil
\leftmark\hbox{}}\def\@evenfoot{}
\def\sectionmark##1{}\def\subsectionmark##1{}}

\oddsidemargin=\evensidemargin
\newcounter{sectionc}\newcounter{subsectionc}\newcounter{subsubsectionc}
\renewcommand{\section}[1] {\vspace{12pt}\addtocounter{sectionc}{1}
\setcounter{subsectionc}{0}\setcounter{subsubsectionc}{0}\noindent
	{\tenbf\thesectionc. #1}\par\vspace{5pt}}
\renewcommand{\subsection}[1] {\vspace{12pt}\addtocounter{subsectionc}{1}
	\setcounter{subsubsectionc}{0}\noindent
	{\bf\thesectionc.\thesubsectionc. {\kern1pt \bfit #1}}\par\vspace{5pt}}
\renewcommand{\subsubsection}[1] {\vspace{12pt}\addtocounter{subsubsectionc}{1}
	\noindent{\tenrm\thesectionc.\thesubsectionc.\thesubsubsectionc.
	{\kern1pt \tenit #1}}\par\vspace{5pt}}

\newcommand{\textlineskip}{\baselineskip=13pt}
\newcommand{\smalllineskip}{\baselineskip=10pt}

\def\eightcirc{
\begin{picture}(0,0)
\put(4.4,1.8){\circle{6.5}}
\end{picture}}
\def\eightcopyright{\eightcirc\kern2.7pt\hbox{\eightrm c}}

\newcommand{\copyrightheading}[1]
	{\vspace*{-2.5cm}\smalllineskip{\flushleft
      {\footnotesize Mod. Phys. Lett. A, submitted for publication   #1}\\
       {\footnotesize $\eightcopyright$\, World Scientific Publishing Company
        }\\
	 }}

\def\abstracts#1#2#3{{
	\centering{\begin{minipage}{4.5in}\baselineskip=10pt\footnotesize
	\parindent=0pt #1\par
	\parindent=15pt #2\par
	\parindent=15pt #3
	\end{minipage}}\par}}

\renewenvironment{thebibliography}[1]
	{\frenchspacing
	 \ninerm\baselineskip=11pt
	 \begin{list}{\arabic{enumi}.}
        {\usecounter{enumi}\setlength{\parsep}{0pt}
	 \setlength{\leftmargin 12.7pt}{\rightmargin 0pt} 
         \setlength{\itemsep}{0pt} \settowidth
	{\labelwidth}{#1.}\sloppy}}{\end{list}}

\newcounter{itemlistc}
\newcounter{romanlistc}
\newcounter{alphlistc}
\newcounter{arabiclistc}

\def\@citex[#1]#2{\if@filesw\immediate\write\@auxout
	{\string\citation{#2}}\fi
\def\@citea{}\@cite{\@for\@citeb:=#2\do
	{\@citea\def\@citea{,}\@ifundefined
	{b@\@citeb}{{\bf ?}\@warning
	{Citation `\@citeb' on page \thepage \space undefined}}
	{\csname b@\@citeb\endcsname}}}{#1}}

\newif\if@cghi
\def\cite{\@cghitrue\@ifnextchar [{\@tempswatrue
	\@citex}{\@tempswafalse\@citex[]}}
\def\citelow{\@cghifalse\@ifnextchar [{\@tempswatrue
	\@citex}{\@tempswafalse\@citex[]}}
\def\@cite#1#2{{$\null^{#1}$\if@tempswa\typeout
	{IJCGA warning: optional citation argument
	ignored: `#2'} \fi}}

\def\@refcitex[#1]#2{\if@filesw\immediate\write\@auxout
	{\string\citation{#2}}\fi
\def\@citea{}\@refcite{\@for\@citeb:=#2\do
	{\@citea\def\@citea{, }\@ifundefined
	{b@\@citeb}{{\bf ?}\@warning
	{Citation `\@citeb' on page \thepage \space undefined}}
	\hbox{\csname b@\@citeb\endcsname}}}{#1}}

\def\@refcite#1#2{{#1\if@tempswa\typeout
        {IJCGA warning: optional citation argument
	ignored: `#2'} \fi}}

\def\refcite{\@ifnextchar[{\@tempswatrue
	\@refcitex}{\@tempswafalse\@refcitex[]}}

\def\pmb#1{\setbox0=\hbox{#1}
	\kern-.025em\copy0\kern-\wd0
	\kern.05em\copy0\kern-\wd0
	\kern-.025em\raise.0433em\box0}

\def\fnt#1#2{\footnotetext{\kern-.3em
	{$^{\mbox{\scriptsize #1}}$}{#2}}}

\def\runninghead#1#2{\pagestyle{myheadings}
\markboth{{\protect\footnotesize\it{\quad #1}}\hfill}
{\hfill{\protect\footnotesize\it{#2\quad}}}}
\font\tenrm=cmr10
\font\tenit=cmti10
\font\tenbf=cmbx10
\font\bfit=cmbxti10 at 10pt
\font\ninerm=cmr9

\font\eightrm=cmr8

\def\qed{\hbox{${\vcenter{\vbox{			
   \hrule height 0.4pt\hbox{\vrule width 0.4pt height 6pt
   \kern5pt\vrule width 0.4pt}\hrule height 0.4pt}}}$}}

\begin{document}

\newpage

\runninghead{Haret C. Rosu}
{KdV barotropic solitons}

\normalsize\textlineskip
\thispagestyle{empty}
\setcounter{page}{1}

\copyrightheading{}    

\vspace*{0.88truein}

\bigskip
\centerline{\bf KdV ADIABATIC INDEX SOLITONS} 
\centerline{\bf IN BAROTROPIC OPEN FRW COSMOLOGIES}
\vspace*{0.035truein}
\vspace*{0.37truein}
\vspace*{10pt}
\centerline{\footnotesize HARET C. ROSU}
\vspace*{0.015truein}
\centerline{\footnotesize  Instituto de F\'{\i}sica,
Universidad de Guanajuato, Apdo Postal E-143, 37150 Le\'on, Gto, Mexico}
\vspace*{0.225truein}

\vspace*{0.21truein}
\abstracts{Applying standard mathematical methods,
it is explicitly shown how the Riccati equation
for the Hubble parameter $H(\eta)$ of
barotropic open FRW cosmologies is connected with a
Korteweg-de Vries equation for adiabatic index solitons.
It is also shown how one can embed a discrete sequence of adiabatic indices
of the type $n^2(\frac{3}{2}\gamma -1)^2$ ($\gamma \neq 2/3$)
in the ${\rm sech}$ FRW adiabatic index soliton.   }{}{}


\textlineskip                  
\vspace*{12pt}                 

\vspace*{1pt}\textlineskip	
\vspace*{-0.5pt}
\noindent


\noindent




\bigskip
\bigskip


{\bf 1} - {\bf Riccati equation for barotropic FRW cosmologies}

\bigskip

\noindent
Recently, Faraoni,\cite{Far} showed that the equations
describing barotropic FRW cosmologies can be combined in a simple
Riccati equation leading in a much easier way
to the textbook solutions for the FRW scale factors.
Faraoni obtained the following cosmological Riccati equation
$$
 \frac{dH}{d\eta}=-cH^2-\kappa c~,
\eqno(1a)
$$
for the log derivative of the FRW scale factor, the famous Hubble
parameter $H(\eta)=\frac{da/d\eta}{a}$.
The independent variable is the conformal time $\eta$, $c$ is simply related
to the adiabatic index of the cosmological fluid,
$c=\frac{3}{2}\gamma -1$ which was assumed constant by Faraoni,
and $\kappa=0,\pm1$ is the curvature index of the flat, closed, open
FRW universe, respectively. For the mathematical considerations of
this work we are more interested in the open cosmologies
$\kappa =- 1$. The flat case is special and has been discussed in a
previous work,\cite{flat} whereas the case $\kappa =+ 1$ will be commented in
subsection 3.2 herein.

We notice that from the mathematical point of view, Faraoni's FRW
Riccati equation being of constant coefficients is directly integrable.
The solutions are
$$
H^+=-\tan (c\eta -\phi _+) ~, \qquad H^-={\rm tanh}(c\eta - \phi _-)~,
\eqno(1b)
$$
for the closed and open FRW universes, respectively. Of course, depending on
the initial conditions, the solutions can be written in terms of the
corresponding cofunctions as well. The minus sign in front of the phases has
been chosen for later convenience.

\bigskip
\newpage

{\bf 2 -} {\bf Cosmological Korteweg de Vries equation}

\bigskip

\noindent
We now pay special attention to the $\kappa =-1$ case for which a connection
with the KdV equation is possible.
Employing the well-known change of function
$H^{-}=\frac{1}{c}\frac{w^{'}}{w}$ one can pass from the nonlinear Riccati
equation to the linear second-order differential equation
$$
w^{''}- c^2w=0~,
\eqno(2)
$$
where the prime means derivation with respect to $\eta$.
This equation can be factorized
$$
A_1A_2w=0
\eqno(3)
$$
by means of the operators
$$
A_1=\left(\frac{d}{d\eta}+ W\right)~,
\qquad \qquad
A_2=\left(\frac{d}{d\eta}-W\right)~,
\eqno(4)
$$
where $W=cH^{-}$.
The general linear solution (the general zero mode) is of the form
$$
w(\eta, T)=\alpha (T) e^{c\eta}+\beta (T) e^{-c\eta}~,
\eqno(5)
$$
where we have assumed that the superposition constants are functions of a
new coordinate $T$ that is independent of $\eta$.
It is easy to show the relationship between the two superposition constants
and the phase $\phi _-$
$$
\phi _{-} (T)=\frac{1}{2}\ln \beta /\alpha~.
\eqno(6)
$$
Inverting as in SUSY QM the order of application of the factorization
operators, one gets an equation of the type
$$
A_2A_1u=0 \qquad \qquad {\rm or} \qquad \qquad u^{''}+c_1^2(\eta , T)u=0~,
\eqno(7)
$$
where $c_1(\eta ,T)$ is the ${\rm sech}$ soliton with the asymptotic wings going
to $c^2$
$$
c_1^2(\eta ,T)=c^2-2c^2{\rm sech} ^2(c\eta -\phi _{-} (T))~.
\eqno(8)
$$
It is now easy to apply basic results from the
KdV mathematics allowing to interpret
$c_1^2$ as a single ${\rm sech}$ soliton of the following KdV
equation,\cite{KdV}
$$
\frac{dC}{dT}-6C\frac{dC}{d\eta}+\frac{d^3C}{d\eta ^3}=0~,
\eqno(9)
$$
if the condition $\frac{d\phi _-}{dT}=4c^3$ is imposed. Moreover, by choosing
initial conditions at $T=0$, one can see that $T$ is the evolution variable
whereas the conformal time plays a role similar to a spatial variable.\cite{fmp}
\bigskip


\bigskip
\bigskip

\newpage

{\bf 3 -} {\bf SUSY recursive procedure}

\bigskip

\noindent
{\bf 3.1 -}
The scheme presented in the previous section can be iterated in a simple and
well-known manner.\cite{r,hr} We put $T={\rm const}$, i.e., we consider an
instantaneous soliton profile
and take $\phi _- =0$ for simplicity. Thus, we start with a
`quantum' mechanical system
in a constant `potential' $c^2$ and relate it
to a Schr\"odinger-type equation in the conformal time domain which has a
fundamental `frequency' at $-c^2$.
We solve the ``fermionic" Riccati equation
i.e.,
$$
W^2_{1}-W^{'}_{1}-c^2=0~,
\eqno(10)
$$
to find Witten's superpotential $W_1(\eta)=-c\tanh [c\eta]$
and next go to the ``bosonic" Riccati equation
$$
W^2_{1}+W^{'}_{1}+c _{1}^{2}(\eta)-c^{2}=0~,
\eqno(11)
$$
in order to get $c _{1}^{2}(\eta)=-2c^{2}
{\rm sech}^2[c\eta]$. Moreover, one can write the Schr\"odinger
equation corresponding to the ``bosonic" Riccati equation as follows
$$
-\tilde{y}^{''}+c _{1}^{2}(\eta)\tilde{y}=-c^{2}\tilde{y}~,
\eqno(12)
$$
with the solution
$\tilde{y}\propto c\,{\rm sech}(c\eta)$.
The physical picture is that of a ${\rm sech}$ soliton profile
containing a
single mode self-trapped at $-c^2$
within the frequency pulse. One can employ the scheme recursively to get
several localized modes embedded in the soliton
profile.  
Indeed, suppose we would like to introduce $N$
adiabatic indices of the type $c _{n}^{2}=-n^2c^{2}$, $n=1,...N$ in the
${\rm sech}$ pulse. Then, one has to solve the sequence of equations
$$
W_{n}^{2}-W_{n}^{'}=c _{n-1}^2+n^{2}c^{2}~,
\eqno(13)
$$
$$
W_{n}^{2}+W_{n}^{'}=c _{n}^2+n^{2}c^{2}~,
\eqno(14)
$$
inductively for $n=1...N$.\cite{r,hr} The soliton
containing $N$ adiabatic indices $n^2c^{2}$, $n=1...N$ is of the form
$c _{N}^{2}(\eta)=-N(N+1)c^{2}{\rm sech}^2(c\eta)$. The
corresponding modes can be written in a compact form as follows
$$
\tilde{y}_{n}(\eta ;N)\approx A^{\dagger}(\eta ;N)A^{\dagger}(\eta ;N-1)
A^{\dagger}(\eta ;N-2)...A^{\dagger}(\eta ;N-n+2){\rm sech} ^{N-n+1}
c\eta~,
\eqno(15)
$$
i.e., by applying the first-order operators
$A^{\dagger}(\eta;a_{n})=-\frac{d}{d\eta}-a_{n}c\tanh (c\eta)$,
where $a_{n}=N-n$, onto the ``ground state" sech mode.



\bigskip

\noindent
{\bf 3.2 -} For the case $\kappa =1$ the ``fermionic"
Riccati equation
$$
W^2_{1}-W^{'}_{1}+c^{2}=0
\eqno(16)
$$
leads to the solution $W_{1}=c\,{\rm tan}(c\eta)$ and from
the ``bosonic" Riccati equation
$$
W^2_{1}+W^{'}_{1}+c _{1}^{2}+c^{2}=0~,
\eqno(17)
$$
one will find $c _{1}^{2}(\eta)
=2c^{2}{\rm sec}^{2}(c\eta)$. Consequently, the Schr\"odinger-like equation
$$
-\tilde{y}^{''}+c _{1}^{2}(\eta)\tilde{y}=c^{2}\tilde{y}
\eqno(18)
$$
has solutions of the type
$\tilde{y}\propto c\,{\rm sec}(c\eta)$, which is not of the localized soliton
type, and therefore the approach leads to unphysical results.


\bigskip



\bigskip

{\bf 4 - Conclusion}

\bigskip

\noindent
A cosmological KdV equation for adiabatic index solitons
in the realm of barotropic FRW models has been
introduced in this work. A simple recursive nonrelativistic SUSY scheme has
been also applied
to show that in the barotropic soliton profile one can embed a discrete
sequence of adiabatic indices. Each eigenmode of the soliton corresponds
to one of the embedded adiabatic indices.

\bigskip
\bigskip
\bigskip
\bigskip

KdVc.tex








\newpage
\bigskip
\noindent
{\bf References}

\begin{thebibliography} {000}
\bibitem{Far} V. Faraoni,
Am. J. Phys. {\bf 67}, 732 (1999) [physics/9901006].
See also, J.A.S. Lima,  Am. J. Phys., to appear, [astro-ph/0109215].
For supersymmetric approach with application to
cosmological acceleration, see H.C. Rosu, Mod. Phys. Lett. A {\bf 15},
979 (2000) [gr-qc/0003108].


\bibitem{flat} H.C. Rosu,  Mod. Phys. Lett. A {\bf 16}, 1147 (2001)
[gr-qc/0003108].

\bibitem{KdV}
It seems that a water KdV sech soliton has been first observed in a
shallow drainage channel during the month of August 1834 by J.S. Russell, who
followed it on horseback in a chase of one or two miles. The KdV equation has
been first obtained in 1895.

\bibitem{fmp}
Using the direction of propagation as an evolution coordinate is a well-known
procedure for pulse propagation in nonlinear materials and in waveguide physics,
see, e.g., G. Fibich, V.M. Malkin, G.C. Papanicolau,
Phys. Rev. A {\bf 52}, 4218 (1995).

\bibitem{r}
J.L. Rosner,
Ann. Phys. {\bf 200}, 101 (1990); W. Kwong and J.L. Rosner, Prog.
Theor. Phys. Suppl. {\bf 86}, 366 (1986).

\bibitem{hr}
H. Rosu, J.L. Romero, J. Socorro, Nuovo Cimento B {\bf 113}, 549 (1998)
[physics/9707012].

\end{thebibliography}
\end{document}